\documentclass[prl,aps,twocolumn,showpacs,groupedaddress]{revtex4}
\usepackage{graphicx}
\begin{document}
\title{Geometric magic numbers of sodium clusters: 
Interpretation of the melting behaviour}

\author{Eva G. Noya}
\affiliation{University Chemical Laboratory, Lensfield Road, Cambridge CB2 1EW,
United Kingdom}
\author{Jonathan P. K. Doye}
\affiliation{University Chemical Laboratory, Lensfield Road, Cambridge CB2 1EW,
United Kingdom}
\author{David J. Wales}
\affiliation{University Chemical Laboratory, Lensfield Road, Cambridge CB2 1EW,
United Kingdom}
\date{\today}

\begin{abstract}

Putative global minima of sodium clusters with up to 380 atoms
have been located for two model interatomic potentials. 
Structures based upon the Mackay icosahedra predominate for both potentials,
and the magic numbers for the Murrell-Mottram model show excellent
agreement with the sizes at which maxima in the latent heat and entropy 
change at melting have been found in experiment.
\end{abstract}

\maketitle

The melting of sodium clusters has been the subject of numerous recent studies.
Much of this interest has arisen due to the availability
of high-quality experimental data, which has
allowed detailed comparisons between theory and experiment. 
In particular, the Haberland group \cite{haberland1,haberland2,haberland3}
have measured the caloric curves
of mass-selected positively-charged sodium clusters with up to 360 atoms,
from which the melting temperature ($T_{melt}$) and latent heat can
be extracted. The values of $T_{melt}$ for these clusters
are on average one third lower than that for bulk, and show
variations of up to $\pm\,$50$\,$K depending on the cluster size. 
There have been a number of theoretical studies that,
using different levels of theory,
have investigated the origins of the size-dependence of $T_{melt}$
\cite{aguado2,calvo1,manninen2,garzon,chacko05}.
However, the peaks in the melting temperature do not seem to correlate
either with the electronic or geometric shell closings of sodium 
clusters, and
none of those theoretical studies have been able to provide a 
satisfactory explanation for the non-monotonic variation of $T_{melt}$.
Significant progress was made in Haberland \emph{et al.}'s most 
recent paper, in which they
observed that the energy and entropy changes on melting 
provide more structural insight into the system than $T_{melt}$
itself \cite{haberland}. 
In particular, these two quantities exhibit pronounced maxima
at certain `magic numbers', 
some of which have a clear interpretation in terms of geometric structures,
such as the Mackay icosahedra, whilst others remain unassigned.
Therefore, a systematic investigation of the geometric structure of 
sodium clusters in this size range would be of great help in the identification
of the structures underlying these magic numbers.

Previous work on the structure of sodium clusters has for the most
part concentrated on clusters with less than 60 atoms 
\cite{BonacicKoutecky91,Rothlisberger91,Calvo00b,Kummel00,Lai02,Solovyov02,Manninen04}.
By contrast, in this Letter we have attempted to locate the lowest-energy 
structures of sodium clusters for all sizes up to $N$=380 using the 
basin-hopping global optimization method \cite{wales}. Such large sizes
necessitate the use of a model potential, and 
we have considered two different forms for the interatomic interactions,
namely the Gupta \cite{gupta,li} and Murrell-Mottram 
(MM) \cite{mm1,mm2,mm3} potentials. 
The MM potential has more parameters, has been fitted to a
wider range of properties, and exhibits good transferability \cite{mm3}.
Therefore, it is expected to be the more reliable of the two potentials, 
but it is also significantly more expensive to compute.
The advantage of considering two potentials is that we can have 
greater confidence in those structural features that are common to both
potentials.

\begin{figure*}
\begin{center}
\includegraphics[width=100mm,angle=-90]{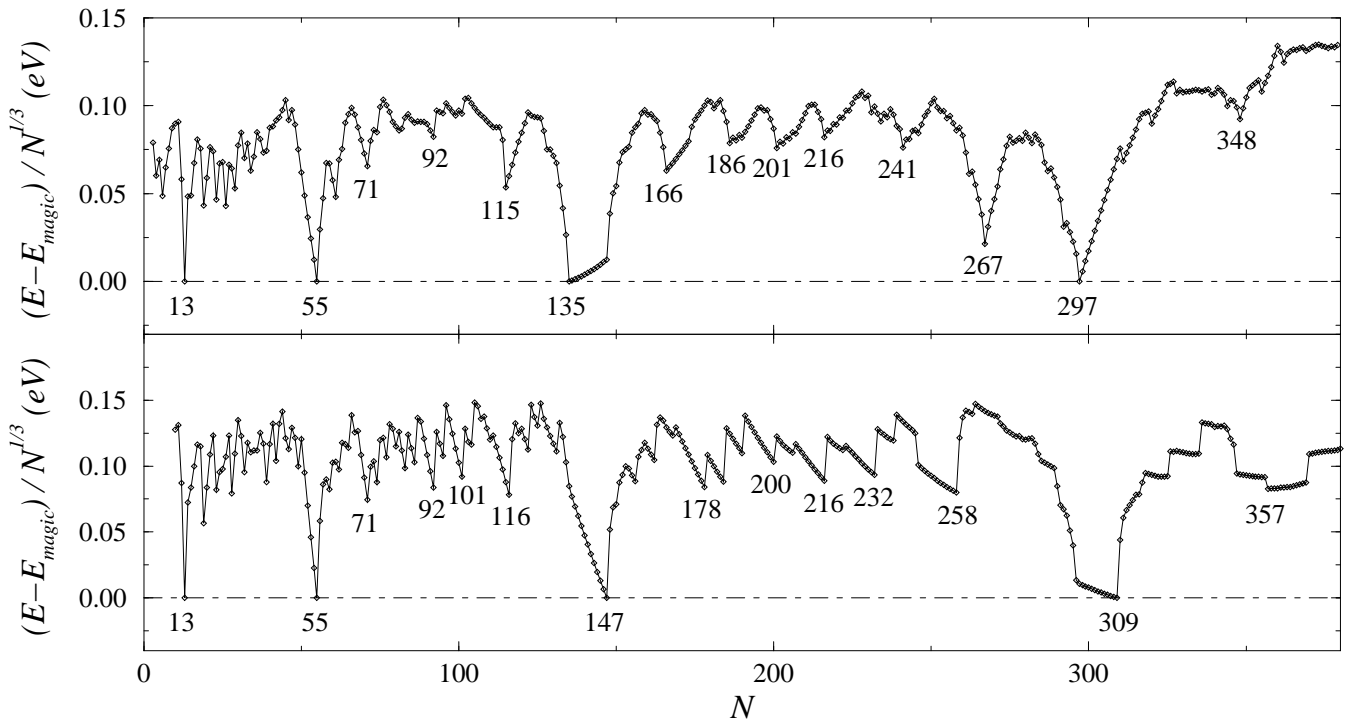}
\caption{\label{fig1} Energy of the global minima found for the
Gupta (upper panel) and MM (lower panel) potentials as a function of
size. Energies are given relative to $E_{magic}$, which is a function fitted
to the energies of the first four stronger magic numbers.
$E_{magic}^{Gupta}=0.0403 - 0.2546 N^{1/3} + 1.2134 N^{2/3}
-1.1568 N$; $E_{magic}^{MM}= -0.4788 + 0.5261 N^{1/3} +
0.9852 N^{2/3} - 1.1110 N$.}
\end{center}
\end{figure*}

\begin{figure*}
\begin{center}
\includegraphics[width=180mm]{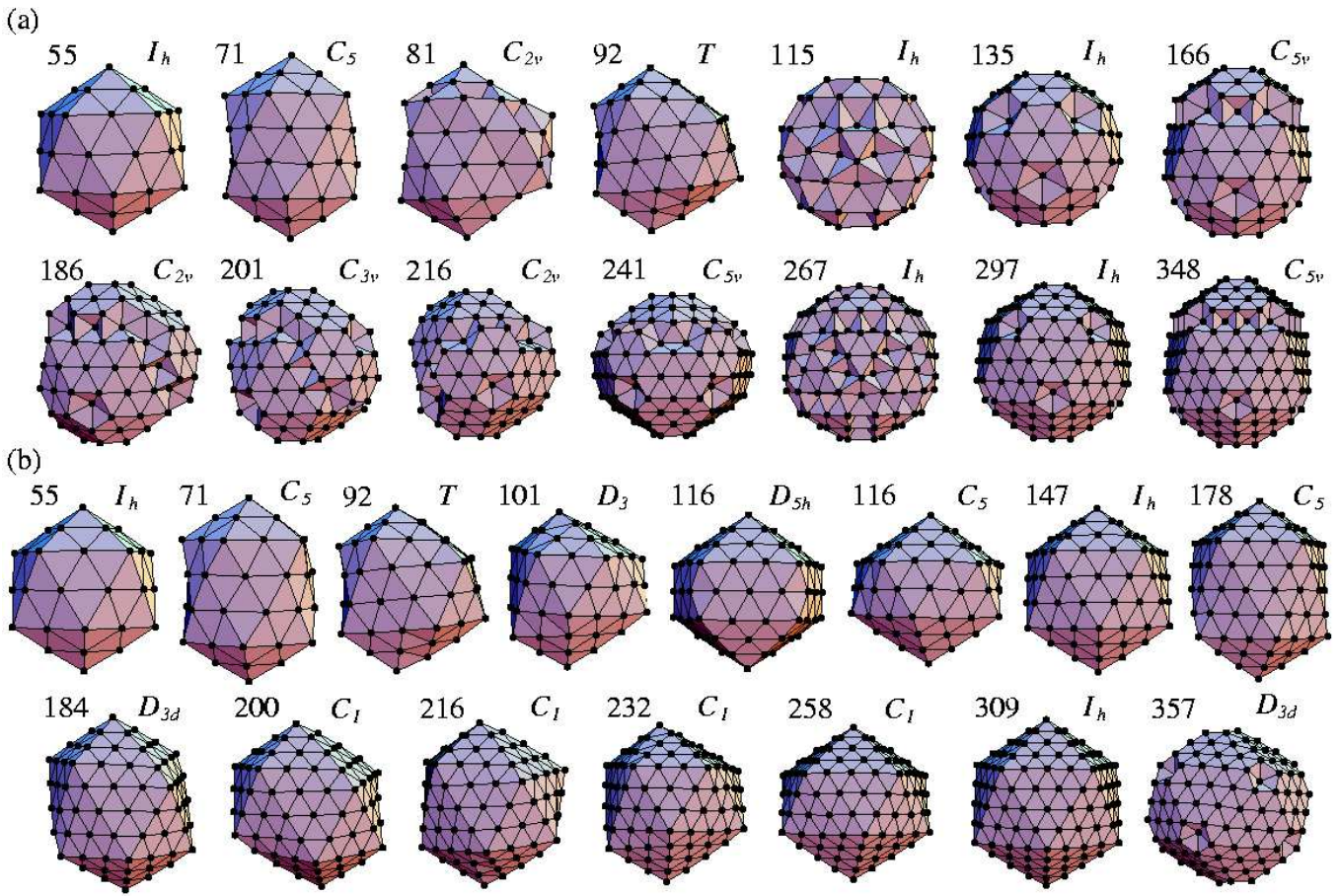}
\caption{\label{fig2} A selection of sodium clusters that show enhanced 
stabilities for the (a) Gupta and (b) Murrell-Mottram potentials.
Each structure is labelled by the number of atoms and its point group symmetry.}
\end{center}
\end{figure*}

In Figure \ref{fig1}, we have plotted the energies of the putative global 
minima for the two potentials, 
and Figure \ref{fig2} shows the structures of some of the 
magic number clusters.
The energies and coordinates for all the structures are available at
the Cambridge Cluster Database \cite{CCD}. 
For $N\le 57$ the Gupta global minima have 
been previously reported by Lai {\it et al.} \cite{Lai02}.

The Haberland group found that for $N<100$ many sodium clusters do not show 
a clear melting transition, but pass from solid to liquid without a pronounced 
latent heat \cite{haberland3}. 
Na$_{55}$ stands in contrast to this trend having a particularly 
high melting temperature, but Na$_{70}$ and Na$_{92}$ also represent exceptions
\cite{haberland}. 
Both potentials exhibit a pronounced magic number at $N$=55, which,
as expected, corresponds to a complete Mackay icosahedron.
Typically, there are two types of overlayer for growth on the surface of a 
Mackay icosahedron.
The first, the Mackay overlayer, continues the face-centred-cubic (fcc) 
packing of the twenty fcc tetrahedra making up the Mackay icosahedron, and 
leads to the next Mackay icosahedron. 
By contrast, the second, the anti-Mackay overlayer, 
adds atoms in sites that are hexagonal close-packed with respect to the 
underlying fcc tetrahedra. Typically, growth starts off in the anti-Mackay 
overlayer because of a greater number of nearest-neighbour interactions, 
but then switches to the Mackay overlayer because it involves less 
strain \cite{Northby,Doye97d}.

Interestingly, structures that do not adopt either of these overlayers are 
prevalent for both potentials. 
The magic number at Na$_{71}$, a possible explanation
for the experimental feature at $N$=70, provides a good example. Both potentials
have the same $C_5$ global minimum, where the five faces around the vertex
of the 55-atom Mackay icosahedron are covered by a Mackay-like cap,
but where both the overlayer and core have been twisted 
with respect to the ideal Mackay sites.
This twist increases the coordination number of some of the surface atoms at
the expense of increased strain and creates a structure where, unlike
both the anti-Mackay and Mackay overlayers, the surface consists entirely of 
$\{111\}$-like faces.
A similar structure is a magic number at $N$=92 and involves the covering
of ten faces with a Mackay overlayer, which then undergoes a twist distortion, 
giving rise to a structure with $T$ point group symmetry, 
instead of $C_{3v}$ for the ideal Mackay geometry. 
These structures look like a hybrid of the 55-atom and 147-atom 
Mackay icosahedra, because they have triangular $\{111\}$ faces of sizes
corresponding to both the smaller and larger Mackay icosahedra.

The Gupta potential does not exclusively exhibit structures based on 
these twisted icosahedra in this size range. For example,
at $N$=81 a structure where eight faces are covered by an 
anti-Mackay overlayer is the global minimum.
This difference between the two potentials becomes more prominent at larger
sizes. For example, there is a feature at $N\approx 116$ in the experimental
results that has been interpreted in terms of a Mackay structure with
15 of the faces of the underlying icosahedron covered \cite{haberland}.
Both potentials have clear features near to this size. For the
MM potential, there is a magic number at $N$=116 and at this size there
are two minima with almost the same energy. The second-lowest minimum 
corresponds to a twisted form of the structure suggested by Haberland 
{\it et al.}, and the lowest-energy isomer is based on a 116-atom 
Ino decahedron but with the central ring of atoms twisted to remove
any $\{100\}$ faces. By contrast, the Gupta potential has a magic number at
$N$=115 that corresponds to an $I_h$ structure with a complete 
anti-Mackay overlayer. This is an unusual feature, since the anti-Mackay
overlayer is usually observed during the initial growth on an icosahedron 
\cite{Northby}, 
but not when that overlayer is nearly complete. 
Moreover, this structure is very high in energy for the MM potential.

Experimentally, Na$_{147}$ is a prominent magic number, 
and, again as expected, the Mackay icosahedron is the global minimum at this
size for both potentials. However, for the Gupta potential a more stable
structure can be obtained by removing the twelve vertex atoms, giving rise
to a magic number at $N$=135 (Fig.\ \ref{fig1}). 
This feature is in clear contradiction with experiment.

For growth on the 147-atom Mackay icosahedron, the differences between
the results for the two potentials become even greater. For the MM potential
structures based upon the twisted icosahedra continue to predominate. However,
the Gupta potential initially exhibits structures with a Mackay overlayer,
and then switches over to an anti-Mackay overlayer near to the completion
of this overlayer at $N$=267.

The MM potential exhibits prominent magic numbers at $N$=178, 216, 232 and 258,
with weaker features at 184, 190, 200, 206, 222 and 238. These structures
correspond to covering successive faces of the 147-Mackay icosahedron
with Mackay-like overlayers, but where the core and surface again undergo
a twist distortion. The 178-, 216-, 232- and 258-atom structures are 
equivalent to the 71-, 92-, 101- and 116-atom structures described above and 
correspond to covering all the faces surrounding 1, 3, 4 and 6 vertices of the
underlying icosahedron. These features are in good agreement with 
the experimental results, which have clear features at $N$=178 and 216,
and a smaller sub-peak at $N$=184. No experimental features have yet been
identified at $N$=232 and 258. However, in this size range the data is sparse,
and the error bars are of similar magnitude to the size variation of the
properties. Therefore, it would be very interesting if further experiments
were conducted at these sizes to examine the predictions of the MM
model.

Interestingly, Haberland {\it et al.}\ suggested undistorted Mackay 
structures to explain the magic numbers at $N$=178 and 216 \cite{haberland}. 
However,
it is more usually found that more stable structures are possible, 
when the five-coordinate atoms at the corners of the added triangular 
faces are not occupied. For example,
this leads to magic numbers at $N$=173 and 213 for Lennard-Jones clusters 
\cite{Romero}. The twist distortion of the icosahedra provides a possible
explanation for this difference in magic numbers. 
As a consequence of the distortion, the coordination number for the corner 
atoms increases from five to six, making it more favourable
for these sites to be occupied.

The magic numbers for the Gupta potential are completely different in this size
range, because of the preference for both undistorted icosahedral structures
and empty vertex sites. The magic numbers at $N$=166, 186, 201, 216 and 241 
are all due to structures with a Mackay overlayer.
If it were favourable for the six-coordinate vertices to be 
occupied, these magic sizes would instead be at $N$=173, 196, 213, 230 and 258.
Only if five-coordinate sites were also occupied would
the magic numbers be 178, 200, 216, 232 and 258.
Analogous to the particular stability of Na$_{115}$ in
the growth of the third shell, there is another magic number
at $N$=267 whose structure involves a complete anti-Mackay layer 
without vertices. 
Closeby ($N=$ 268), Haberland \emph{et al.}\
found a well-structured photoelectron spectrum, 
but they attributed this feature to the existence of an electronic shell
closing rather than to high point group symmetry \cite{haberland}. Furthermore, 
this complete anti-Mackay icosahedra again lies very high in energy for the MM 
potential.

As for the third shell, the complete Mackay icosahedron
is not a magic number for the Gupta potential, but instead an
icosahedron with twelve missing vertices is more stable, 
displacing the magic number to $N$=297. 
The MM potential still predicts the magic number to be at $N$=309, 
but the difference in stability between the 297- and 309-atom structures is 
much smaller. 
Indeed, at $N\approx 360$
structures with missing vertices actually become more stable. 
The similar behaviour of the two potentials suggests that
the loss of vertex atoms is a robust structural feature for sodium;
the two potentials only differ in the size at which this effect first appears.
These results suggest that a plausible explanation of the absence of 
an experimental magic number at $N$=309, but the appearance of a 
feature at $N$=298, is the greater stability 
of a Mackay icosahedron that has lost its vertices. 
However, Haberland \emph{et al.} found that the measured photoelectron
spectrum for Na$_{298}$ is not compatible with such a structure 
\cite{haberland}. 
Furthermore, on measuring the
photoelectron spectrum of Na$_{309}$ as a function of temperature, 
Haberland \emph{et al.} found
that a structural transition occurred at about $40\,$K 
below melting \cite{haberland}. 
Parallel tempering simulations using the Gupta potential, 
however, did not show evidence of any transitions prior to melting 
for Na$_{309}$.
Therefore, our results are unable to offer a structural explanation 
compatible with all the experimental findings associated with the completion
of the fourth icosahedral shell, and their origin remains somewhat mysterious. 

Finally, for growth of the fifth icosahedral shell, the same patterns continue, 
i.e.\ Mackay overlayers for the Gupta potential and twisted icosahedra for 
the MM potential. 
In this size range, experiments predict a peak at $N$=360, 
for which the MM magic number at $N$=357 offers a possible explanation.

In summary, our results support the conclusions derived from Haberland 
{\it et al.}'s recent analysis of the melting behaviour of sodium 
clusters---namely that the clusters in this size range are predominantly 
icosahedral. Of the two potentials we have considered, the structures obtained
for the MM potential appear more reliable and the magic numbers are
in excellent agreement with experiment. In particular, we
suggest that the experimental features at sizes intermediate between
the complete Mackay icosahedra are due to icosahedral structures with
a Mackay overlayer, but where both the core and overlayer undergo
a twist distortion to give structures that have only $\{111\}$-like
faces. It is noteworthy that such features cannot be captured by
a pairwise-additive interatomic potential.

\acknowledgements

The authors are grateful to the Ram\'on Areces Foundation (E.G.N.)
and the Royal Society (J.P.K.D) for financial support. Fruitful
discussions with H.\ Haberland are acknowledged.

\end{document}